\documentclass[aps,prl,twocolumn,showpacs,groupedaddress]{revtex4}
\usepackage{graphicx}
\usepackage{epsfig}
\usepackage{amssymb,amsmath,amsfonts,hyperref}
\usepackage{wasysym}
\usepackage{latexsym}
\usepackage{verbatim}
\usepackage[normalem]{ulem}
\usepackage{color}

\begin{document}

\title{Scaling Theory for the Frictionless Unjamming Transition}
\author{Kabir Ramola} 
\email{kramola@brandeis.edu}
\author{Bulbul Chakraborty}
\email{bulbul@brandeis.edu}
\affiliation{Martin Fisher School of Physics, Brandeis University, Waltham, MA 02454, USA}

\date{\today} 


\begin{abstract}
{We develop a scaling theory of the unjamming transition of soft frictionless disks in two dimensions by defining  local areas, which can be uniquely assigned to each contact.  These serve to define {\it local} order parameters, whose distribution exhibits divergences as the unjamming transition is approached.  We derive scaling forms for these divergences from a mean-field approach that treats the local areas as non interacting entities, and demonstrate that these results agree remarkably well with numerical simulations.  We find that  the asymptotic behaviour of the scaling functions arises from the geometrical structure of the packing while the overall scaling with the compression energy depends on the force law. We use the scaling forms of the distributions to determine the scaling  of the total grain area $A_G$, and the total number of contacts $N_C$.}
\end{abstract}

\maketitle

{\it Introduction:}
The jamming of soft particles has been used as a paradigmatic model of granular\cite{van_hecke_review_2010,liu_nagel_condmat_2010,bolton_weaire_prl_1990,
durian_prl_1997,brujic_makse_physicaa_2003, dinsmore_science_2006,
nordstrom_prl_2010,mailman_prl_2009,lerner_pnas_2012}
and glassy systems \cite{cipelletti_jpcm_2005,ikeda_prl_2012,seth_nature_mat_2011}, active matter \cite{henkes_pre_2011} and biological tissues \cite{bi_nat_mat_2015}. Frictionless soft disks and spheres serve as a first approximation to many theoretical models and have been extensively investigated over the last decade \cite{makse_prl_2000,ohern_prl_2002,silbert_grest_landry_pre_2002,
ohern_pre_2003,wyart_thesis,wyart_epl_2005,silbert_liu_nagel_prl_2005,
henkes_chakraborty_ohern_prl_2007,henkes_chakraborty_pre_2009,
ellenbroek_pre_2009,wyart_prl_2012,
goodrich_liu_sethna_pnas_2016,
ramola_chakraborty_arxiv_2016}.
The unjamming transition of soft spheres exhibits properties reminiscent of critical points in equilibrium systems. Observations include power laws \cite{ohern_prl_2002}, a scaling form for the energy analogous to free energy and resulting relationships between scaling exponents \cite{goodrich_liu_sethna_pnas_2016}, scaling collapse of dynamical quantities such as viscosity \cite{olson_teitel_prl_2007}, and indications of diverging length scales \cite{silbert_liu_nagel_prl_2005}.
Many scaling properties of soft particles near the jamming transition have been analysed in detail \cite{ellenbroek_prl_2006,goodrich_prl_2012}, and finite-size scaling studies seem to suggest a mixed order transition with two critical exponents \cite{ohern_pre_2003,silbert_liu_nagel_prl_2005}. 

Despite considerable effort towards a unifying theory, a clear description of unjamming is still lacking, and the origin of various power laws in this system have remained somewhat mysterious.
Theories so far have focussed on the behavior of global quantities such as energy, packing fraction, pressure, stresses, and the total contact numbers. This is in contrast to the norm in studying critical points where a local order parameter and its distribution within the system is of primary importance. In this Letter we highlight the emergence of diverging contributions to distributions of  local quantities, and show how the underlying disorder of the contact network naturally lead to these divergences.
This in turn leads to non-trivial power laws involving global quantities such as the excess contact number, and the areas occupied by grains. 

\begin{figure}
\includegraphics[width=1\columnwidth,angle=0]{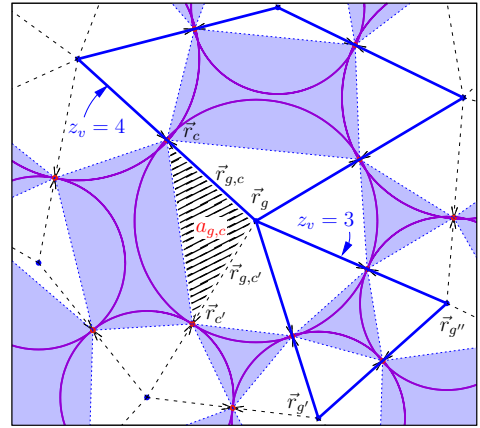}
\caption{A section of a jammed configuration of soft frictionless disks. The centers of the grains with radii $\{\sigma_g\}$ are located at positions $\{ \vec{r}_{g} \}$. 
The contact points between grains are located at positions $\{ \vec{r}_{c} \}$, with contact vectors $\vec{r}_{g,c} = \vec{r}_{c} - \vec{r}_{g}$. The distance vectors $\vec{r}_{g,g'} = \vec{r}_{g'} - \vec{r}_{g}$ form a network of faces (minimum cycles) with $z_v$ sides each.
The polygonal tiling associated with the packing partitions the space into areas occupied by grains (white) and areas occupied by voids (blue).
The triangle formed by the points $(\vec{r}_{g}, \vec{r}_{c}, \vec{r}_{c'})$ (shaded area) is uniquely assigned to the contact $c$ and has an associated area $a \equiv a_{g,c}$, with a normalized area $\alpha_{c} = a_{g,c}/\sigma_g^2$.}
\label{vector_labels_fig}
\end{figure}

Our treatment relies on assigning local grain areas to triangular units uniquely associated with individual contacts, which play the role of ``quasiparticles".
We use properties of the underlying distribution of interparticle distances and angles to derive a probability distribution of these areas, and compare these predictions to results of numerical simulations.
As will be clear from our analysis, the appearance of  triangular units as the basic objects in the scaling theory highlights the importance of {\it three-body terms} as opposed to two-body terms such as interparticle distances that have been considered in the literature.

We focus specifically on the unjamming transition of soft disks, i.e. we approach the transition point from mechanically stable (jammed) packings with decreasing energies ($E_G \to 0^{+}$). 
In such jammed states, the disks organize into  complicated ``random" structures which are hard to characterise owing to the complexity of the non-convex curved shapes formed by voids. In order to avoid this problem we construct polygonal tilings that partitions space into areas occupied by grains and areas occupied by the voids (see Fig. \ref{vector_labels_fig}).  This construction\cite{ramola_chakraborty_arxiv_2016} bears similarities to the ``quadron" framework \cite{blumenfeld_epje_2006, blumenfeld_granular_2012, blumenfeld_prl_2014}. 

%

\begin{figure}
\includegraphics[width=1\columnwidth,angle=0]{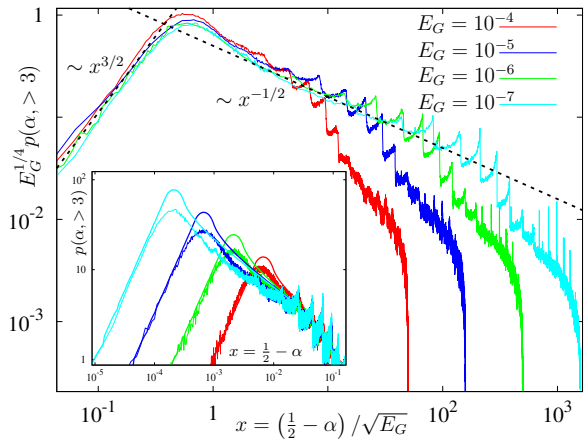}
\caption{Scaling collapse of the distribution of areas $p(\alpha,>3)$ of the $z_v >3$ cycles at different energies. The plot shows distributions for $N_G = 4096$ disks interacting via harmonic potentials ($\mu = 2$). $x \to 0$ corresponds to disks with contact angles close to $\pi/2$.
The scaling is consistent with Eq. (\ref{scaling_ansatz_disorder}). The limiting behaviours of the distribution are provided in Eq. (\ref{scaling_function_disorder}). {\bf (Inset)} Comparision between the distributions obtained from the theory (bold lines) and numerical simulations demonstrating very good agreement.}
\label{scaling_collapse_disordered_figure}
\end{figure}

We then assign the polygonal grain areas to triangular units $\alpha_c$ (normalized by the size of each disk), uniquely associated with each contact $c$. This defines a reliable local order parameter for the unjamming transition \cite{ramola_chakraborty_arxiv_2016}.
The probability distribution of these areas displays divergences at well defined values of $\alpha$ that become sharper as the transition is approached. We identify these as arising from specific structures within the jammed state. The distribution of areas is best expressed as 
\begin{equation}
p(\alpha) = \underbrace{p_{\rm reg}(\alpha) + p_{\rm DO}(\alpha)}_{p(\alpha,> 3)} + \underbrace{p_{\rm O}(\alpha)}_{p(\alpha,3)},
\label{palpha_distribution}
\end{equation}
where $p_{\rm DO}$ and $p_{\rm O}$ are classified as ``disordered" and ``ordered" divergences respectively. Disordered divergences arise from cycles (see Fig. \ref{vector_labels_fig}) with four or more disks in contact ($z_v> 3$, labelled as $>3$ for brevity), and the ``ordered'' ones arise within cycles formed by three disks ($z_v = 3$, labelled as $3$). $p_{\rm reg}$ represents the regular part of the distribution that does not have a diverging energy dependence. 
The main result of this Letter is the derivation, and verification through numerical simulations of a scaling form for $p_{\rm DO}$ (Fig. \ref{scaling_collapse_disordered_figure}), which displays a divergence at $\alpha = 1/2$, 
\begin{equation}
p_{\rm DO}(\alpha) = {E_G}^{-1/2\mu} \mathcal{P}_{\rm DO} \left(\frac{\frac{1}{2}-\alpha}{{E_G}^{1/\mu}} \right)~,
\label{scaling_ansatz_disorder}
\end{equation}
where $\mu$ characterizes the interparticle potential ($\mu =2$ for harmonic potentials).
The scaling function possesses the following asymptotic behaviour:
\begin{eqnarray}
\mathcal{P}_{\rm DO}(x) \sim
\begin{cases}
x^{3/2} \;, \; x \to 0, \\
\\
x^{-1/2} \;, \; x \to \infty \;.
\end{cases}
\label{scaling_function_disorder}
\end{eqnarray}
Similarly, the ``ordered" divergence has a scaling form
\begin{equation}
p_{\rm O}(\alpha) = {E_G}^{-1/\mu} \mathcal{P}_{\rm O} \left(\frac{\frac{\sqrt{3}}{4}-\alpha}{{E_G}^{1/\mu}} \right),
\label{scaling_ansatz_order}
\end{equation}
which is integrable in the $E_G \to 0^{+}$ limit.
The scaling functions do not depend on the interaction potential.

The divergences in $p(\alpha)$ are reminiscent of van Hove singularities in the vibrational density of states in crystals \cite{VanHove} which are broadened by thermal disorder.  
The divergences in  $p(\alpha)$  are broadened at finite $E_G$, becoming infinitely sharp only as $E_G \rightarrow 0^+$.  These singular distributions are in sharp contrast to the broadening of order parameter distributions approaching a thermal critical point.   
We will show that the power laws describing the evolution of global quantities approaching  unjamming are a consequence of the singularities of  $p_{\rm DO}(\alpha)$.  
%
In particular, the total number of contacts ($N_C$), with $\Delta N_C = N_C(E_G) - N_C(0)$ scales as:
\begin{equation}
\Delta N_C \sim E_G^{1/2\mu},
\label{scaling_z_vs_E}
\end{equation}
a form observed in several studies of jamming\cite{ohern_prl_2002,wyart_thesis,van_deen_pre_2014,goodrich_liu_sethna_pnas_2016,
ramola_chakraborty_arxiv_2016}.
The scaling of the total grain area $A_G$, with $\Delta A_G = A_G(E_G) - A_G(0)$,  follows the scaling of $\Delta N_C$.

{\it Energy Ensemble and Local Areas:}
We perform our analysis in a fixed energy-volume ensemble $(E_G,V)$
\cite{ramola_chakraborty_arxiv_2016} of jammed states of soft frictionless disks in two-dimensions.   The microstates of this ensemble are specified by grain positions $\{\vec{r}_{g}\}$ and radii $\{\sigma_{g}\}$ that yield a force balanced state at a given energy $E_G$.  We keep the volume of the total space fixed ($L_x = L_y = 1$). We consider disks interacting via a repulsive potential
\begin{equation}
V[\{\vec{r_g},\sigma_{g}\}] = \sum_{g \ne g'}\frac{1}{\mu} \left(1-\frac{|\vec{r}_{g,g'}|}{\sigma_{g,g'}}\right)^\mu \Theta\left(1-\frac{|\vec{r}_{g,g'}|}{\sigma_{g,g'}}\right),
\label{potential_energy_functional}
\end{equation}
with $\mu > 1$, $\vec{r}_{g,g'} = \vec{r}_{g'} - \vec{r}_{g}$, $\sigma_{g,g'} = \sigma_{g} + \sigma_{g'}$, 
and the energy of a microstate is $E_G = \sum_{g} V[\{\vec{r_g},\sigma_{g}\}]$.

Each jammed state of frictionless disks is characterised by a system spanning contact network which naturally partitions the space into convex minimum cycles (or faces) of $z_v$ sides each (see Fig. \ref{vector_labels_fig}).
The system can then be parametrized in terms of the interparticle distance vectors $\{\vec{r}^{i}_{g,g'}\}$ where the index $i$ labels the vectors within each cycle.  The  loop constraints around each face $\sum_{i} \vec{r}^{i}_{g,g'} = 0$, account for the overcounting of the degrees of freedom.
As we show \cite{SI}, these constraints provide the crucial correlations that determine the internal structures and in turn the scaling behaviour near the unjamming transition.
The positions of the contacts are represented by  $\{ \vec{r}_{c} \}$ with $\vec{r}_{c} = \vec{r}_g + \frac{\sigma_g}{\sigma_g + \sigma_g'}\left( \vec{r}_{g'} - \vec{r}_{g} \right)$, where $c$ is the contact between grains $g$ and $g'$, and contact vectors $\vec{r}_{g,c} = \vec{r}_c - \vec{r}_g$. Each contact is counted twice, once for each grain (see Fig. \ref{vector_labels_fig}).
Following the network representation introduced in \cite{ramola_chakraborty_arxiv_2016}, we define local and global order parameters, respectively,  as the areas:
\begin{eqnarray}
a_{g,c} = \frac{1}{2}(\vec{r}_{g,c} \times \vec{r}_{g,c'}) ~~{\rm and}~~ A_G = \sum_{c = 1}^{N_C} a_{g,c}.
\label{grain_area_eq}
\end{eqnarray}
where $\vec{r}_{g,c}$ and $\vec{r}_{g,c'}$ are adjacent contact vectors (see Fig. \ref{vector_labels_fig})
and the convention is that the area bounded by $(c,c')$ is uniquely assigned to the contact $c$. 
These individual areas, $a_{g,c}$,  which  play the role of a local packing fraction in our description  can vary between $0$ and $\frac{1}{2} {\sigma_g}^2$ where $\sigma_g$ is the radius of grain $g$. 





{\it Distribution of areas:}
We begin by deriving the scaling behavior of the distribution of areas based on some simplifying assumptions, and then compare the derived results to ones observed in numerical simulations.  We assume that (i) the underlying system is {\it disordered} and has reproducible local distributions, (ii) the distribution of contact vectors  is independent of their orientation, and (iii) there are no correlations between the contact triangles beyond those required by the loop constraints \cite{SI}.  The comparison to numerical simulations demonstrates that this mean-field theory provides an accurate description of the  scaling forms.
In order to account for the varying sizes of the grains between configurations at a given $E_G$, we work with the normalized area,  $\alpha_c = a_{g,c}/{\sigma_g}^2 $, which is bounded between $[0,\frac{1}{2}]$. Similarly, we normalize the contact vectors by the size of the disks, with $|\vec{r}_{g,c}| \to |\vec{r}_{g,c}|/\sigma_g$ (to avoid a proliferation of symbols) now being bounded between $[0,1]$.

In a disordered jammed state, the overlaps between disks $\Delta r_{g,c}$ with
$|\vec{r}_{g,c}| = 1 - \Delta r_{g,c}$, vary between contacts and can be treated as random variables with a reproducible distribution $p(\Delta r_{g,c})$ depending on $E_G$. Using $E_G = \frac{1}{N_G}\sum_{i = 1}^{N_C} (\Delta r_{g,c})^\mu$ (Eq. (\ref{potential_energy_functional})), naturally leads to the following scaling form for the distribution of overlaps
\begin{equation}
p(\Delta r_{g,c}) = \frac{1}{E_G^{1/\mu}} \mathcal{P}_r\left( \frac{\Delta r_{g,c}}{E_G^{1/\mu}} \right).
\label{contact_vector_scaling}
\end{equation}
Although the contact vectors, $\vec{r}_{g,c}$,  have a complicated joint distribution, we focus on $p(\vec{r}_1,\vec{r}_2)$, which is the joint probability of occurrence of contact vectors $\vec{r}_1$, $\vec{r}_2$ at two contiguous edges of a minimum cycle,  bounding a given area $\alpha$. 
The probability of each individual area is then
\begin{small}
\begin{equation}
p(\alpha) = \int \vec{dr_1} \int \vec{dr_2} ~p(\vec{r}_1,\vec{r}_2)  ~\delta\left( \frac{1}{2}|\vec{r}_1||\vec{r}_2| \sin \theta  - \alpha \right) ~,
\label{basic_area_equation}
\end{equation}
\end{small}
where $\theta$ is the relative angle between the two vectors.
We can next express the joint distribution as
\begin{equation}
p(\vec{r}_1,\vec{r}_2) = p(|\vec{r}_1|) p(|\vec{r}_2|) \rho(\theta) ~,
\label{two_point_distribution}
\end{equation}  
with $p(\vec{r}_1) = \int d^2 \vec{r}_2 p(\vec{r}_1, \vec{r}_2) = \frac{1}{2 \pi} p(|\vec{r}_1|)$.
In Eq. (\ref{two_point_distribution}), we have extracted the overall scaling with energy into the first two terms involving the magnitudes, encoding the correlations  in  $\rho(\theta)$. As detailed in \cite{SI}, we treat these correlations within a mean-field framework that incorporates the loop constraints on the contact vectors exactly. A systematic diagrammatic expansion \cite{SI} shows that $\rho(\theta)$  and consequently $p(\vec{r}_1,\vec{r}_2)$ has different behaviours within cycles with $z_v > 3$ and $z_v = 3$.  Importantly, cycles with $z_v >3$ contribute a finite amount to $\rho(\theta)$ at $\theta=\pi/2$ whereas $z_v = 3$ do not.


{\it Scaling forms:}
From Eq. (\ref{basic_area_equation}), it is clear that if the lengths of the contact vectors are held fixed, the vanishing slope of the sine function leads to a singularity in $p(\alpha)$ at $\theta = \pi/2$ (analogous to van Hove singularities arising from vanishing gradients).  As $E_G \to 0^+$, the fluctuations in $\Delta r_{g,c}$ decrease (Eq. (\ref{contact_vector_scaling})), leading to a sharpening divergence.
To proceed, we split the area distribution for $z_v > 3$ into a divergent part $p_{\rm DO}$ arising from angles close to $\pi/2$, and a regular part $p_{\rm reg}$ that arises from the rest
\begin{equation}
p(\alpha,>3) = p_{\rm reg}(\alpha) + p_{\rm DO}(\alpha).
\label{split_distribution}
\end{equation}
Without loss of generality,  we assume that $\rho(\theta,>3)$ near $\theta = \pi/2$  contributing to $p_{\rm DO}(\alpha)$, can be represented as a uniform distribution, $\rho_{\pi/2}$ in the range $[\frac{\pi}{2}-\mathcal{E},\frac{\pi}{2} + \mathcal{E}]$, the corrections are of higher order in $\mathcal{E}$. Then integrating Eq. (\ref{palpha_distribution}) over the full range of $\alpha$ leads to the normalization
\begin{equation}
\int_{0}^{1/2} p_{\rm reg}(\alpha) d \alpha = 1 - \int_{0}^{1/2} p_{\rm O}(\alpha) d \alpha - 2 \mathcal{E} \rho_{\pi/2}.
\label{regular_normalization}
\end{equation}
Since $p_{\rm O}(\alpha)$ is integrable (Eq. (\ref{scaling_ansatz_order})), the only energy dependence of $p_{\rm reg}(\alpha)$ arises from the width $\mathcal{E}$.
To derive $p_{\rm DO}(\alpha)$, we change variables $\{\theta \to \sin \theta\}$ giving
\begin{small}
\begin{equation}
\rho(\sin \theta,>3) = \rho_{\pi/2}\left(1-\sin^2 \theta\right)^{-1/2}  ~~~\left|\theta- \frac{\pi}{2}\right| < \mathcal{E}.
\end{equation}
\end{small}
Next, performing the integration over $\sin \theta$ in Eq. (\ref{basic_area_equation}) using the above expression leads to:
\begin{equation}
p_{\rm DO}(\alpha) = \rho_{\pi/2} \int_{0}^{1} dr_1 \int_{0}^{1} dr_2
~ \frac{p(r_1) p(r_2)}{\sqrt{r_1^2 r_2^2 - 4 \alpha^2}}  G(r_1,r_2,\alpha),
\label{disordered_integral}
\end{equation}
where $G(r_1,r_2,\alpha)$ is a product of theta functions that ensures $\sin \left(\frac{\pi}{2} - \mathcal{E}\right) < \frac{2 \alpha}{r_1 r_2} < 1$.
Although the integral in Eq. (\ref{disordered_integral}) does not have a simple closed form answer for general $p(r)$, it is clear that $p_{\rm DO}(\alpha) $ has a singularity as $\alpha \to 1/2$ and as $r_1 \to 1$ and $r_2 \to 1$, and it is straightforward to extract the scaling behaviour announced in Eq. (\ref{scaling_ansatz_disorder}).
In order to compute the scaling function, we replace the distribution of the contact vectors in Eq. (\ref{contact_vector_scaling}) with a uniform distribution, allowing us to perform the integration exactly.
As shown in \cite{SI},  the scaling form announced in Eq. (\ref{scaling_function_disorder}) follows.   From this analysis, it is evident that the exponents $1/2$ and $3/2$ appearing in  the scaling function (Eq. (\ref{scaling_function_disorder})) arise from the purely geometric nature of the divergence at $\theta \simeq \pi/2$, whereas the scaling with $E_G$ is a consequence of the scaling of the distribution of contact lengths and is controlled by the force law.
As shown in \cite{SI}, the distribution of angles for the $z_v = 3$ cycles are centered around a finite value $\theta = \arcsin{\frac{\sqrt{3}}{2}}$. This leads to an integrable divergence in the distribution of areas from Eq. (\ref{basic_area_equation}) as  $E_G \to 0^+$, and the scaling form  announced in Eq. (\ref{scaling_ansatz_order}) follows.
The contribution from these ordered structures to the disordered divergence at $\theta = \pi/2$ is therefore exponentially suppressed.

{\it Numerical Simulations:} 
In order to test the predictions made by our theory, we perform numerical simulations for a system of  bidispersed disks with diameter ratio $1:1.4$ interacting via harmonic potentials ($\mu = 2$). Configurations are produced using a variant of the O'Hern protocol \cite{ohern_prl_2002}. The energies simulated range from $E_G = 10^{-15}$ to $10^{-3}$, with the number of disks ranging up to $N_G = 8192$.
A scaling collapse of the distributions according to the scaling form in Eq. (\ref{scaling_ansatz_disorder}) is illustrated in Fig. \ref{scaling_collapse_disordered_figure} along with the two limiting behaviours announced in Eq. (\ref{scaling_function_disorder}). The inset of Fig. \ref{scaling_collapse_disordered_figure} illustrates the remarkable agreement between the theoretical distributions and the ones obtained  from numerical simulations.

\begin{figure}
\includegraphics[width=1\columnwidth,angle=0]{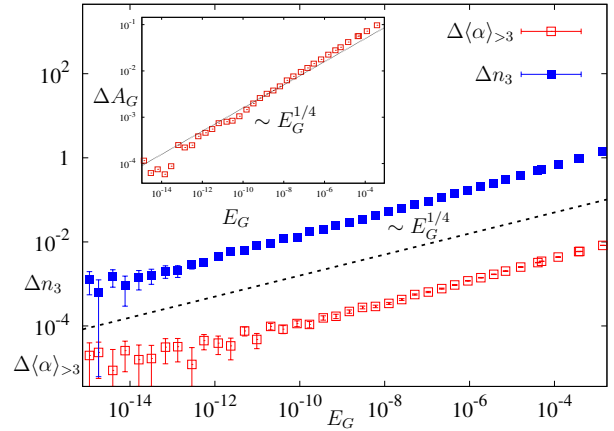}
\caption{Scaling of global quantities with energy for the same system as Fig. \ref{scaling_collapse_disordered_figure}: (i) the excess number of contacts in the $z_v =3$ cycles ($\Delta n_3 = n_3(E_G)- n_3(0)$) (ii) the excess normalized area per contact of the $z_v > 3$ cycles ($ \Delta \langle \alpha \rangle_{>3} = \langle \alpha \rangle_{>3}(E_G) - \langle \alpha \rangle_{>3}(0)$). The scaling is consistent with predictions in Eqs. (\ref{number_of_contacts_scaling}) and (\ref{average_areas_scaling})
{\bf (Inset)} Scaling of excess grain area $\Delta A_G = A_G(E_G) - A_G(0)$ displaying a scaling consistent with $\Delta A_G \sim \Delta N_C$ and Eq. (\ref{scaling_z_vs_E}).}
\label{scaling_of_cycles_fig}
\end{figure}

{\it Scaling of Global Quantities:}
We can use the scaling with $E_G$ of $p(\alpha)$ to derive global scaling properties of the system as the unjamming transition is approached.
Since the microscopic areas are uniquely assigned to a contact, the incremental global area covered by  grains scales as $\Delta A_G \sim \Delta N_C$.
To connect $p(\alpha)$ to the number of contacts $N_C$, we define $g(\alpha)$, the density of states of normalized areas, which we split in a manner similar to $p(\alpha)$ in Eq.(\ref{palpha_distribution}) as
\begin{equation}
g(\alpha) = N_Cp(\alpha) = \underbrace{N_C p_{\rm DO}(\alpha)}_{=g_{\rm DO}(\alpha)} + \underbrace{N_C p_{\rm reg}(\alpha)}_{=g_{\rm reg}(\alpha)} +  \underbrace{N_C p_{\rm O}(\alpha)}_{=g_{\rm O}(\alpha)}.
\label{density_to_probability}
\end{equation} 
The regular part $g_{\rm reg}(\alpha)$ represents the density of areas away from the  divergences and is independent of $E_G$. 
However $p_{\rm reg}(\alpha)$ has an energy dependence from the normalization (Eq. (\ref{regular_normalization})). To extract this dependence, we need to fix $\mathcal{E}$ in a self-consistent manner. 
The height of the peak of $p_{\rm DO}(\alpha)$ scales as $E_G^{-1/2 \mu}$, while the width scales as $E_G^{1/\mu}$ (Eq. (\ref{scaling_ansatz_disorder})). 
The contribution from $p_{\rm DO}(\alpha)$ to the normalization in Eq. (\ref{regular_normalization}) therefore scales as $E_G^{1/2 \mu}$, leading to 
\begin{equation}
2 \mathcal{E} \rho_{\pi/2}  \sim  E_G^{1/2 \mu}.
\label{delta_scaling}
\end{equation}
Then using Eq. (\ref{density_to_probability}) corresponding to the regular part, and the normalization in Eqs. (\ref{regular_normalization}) and (\ref{delta_scaling}), we obtain
\begin{equation}
N_C (E_G)  = \frac{\int_{0}^{1/2}g_{\rm reg}(\alpha) d\alpha}{\int_{0}^{1/2}p_{\rm reg}(\alpha) d\alpha} \approx  N_{C}(0) + N_{C,1/2 \mu} E_G^{1/2 \mu} + ...
\label{NC_scaling}
\end{equation}
which is the scaling relation mentioned in Eq. (\ref{scaling_z_vs_E}). In the inset of Fig. \ref{scaling_of_cycles_fig} we show the scaling of $\Delta A_G$ with energy, which displays a scaling consistent with $\Delta A_G \sim \Delta N_C$ and Eq. (\ref{scaling_z_vs_E}).
Two {\it new} predictions also emerge from a more detailed consideration of divergences in the area distributions \cite{ramola_chakraborty_unpublished}. Defining $n_3$ and $n_{>3}$ as the total number of contacts in cycles with $z_v = 3$ and $z_v >3$ respectively, the excess number of contacts in different cycles ($\Delta n_{\ge 3} = n_{\ge 3}(E_G)- n_{\ge 3}(0)$) scale as
\begin{equation}
\Delta n_3 \sim E_G^{1/2 \mu} ; ~~~\Delta n_{>3} \sim \mathcal{O}(E_G^{1/\mu}).
\label{number_of_contacts_scaling}
\end{equation}
Defining $\langle \alpha \rangle_{3}$ and $\langle \alpha \rangle_{>3}$ as the normalized areas per contact in the different cycles, $ \Delta \langle \alpha \rangle_{\ge 3} = \langle \alpha \rangle_{\ge 3}(E_G) - \langle \alpha \rangle_{\ge 3}(0)$ scales with energy as
\begin{equation}
\Delta \langle \alpha \rangle_3 \sim \mathcal{O}(E_G^{1/\mu}); ~~~\Delta \langle \alpha \rangle_{>3} \sim E_G^{1/2 \mu}.
\label{average_areas_scaling}
\end{equation}
The observed scaling of these global quantities for harmonic potentials is compared with predictions in Fig. \ref{scaling_of_cycles_fig}.

{\it Discussion:}
We identified local units of areas associated with contacts as an order parameter associated with the unjamming transition.  
The marginal state at unjamming is characterized by singularities in the distribution of these local areas. The primary scaling in the system arises from contact vectors with relative angles close $\pi/2$, which lead to a high susceptibility of these contact areas to changes in the compression energy.  This large susceptibility, which is a signature of the marginal state, is reminiscent of  van Hove singularities that render crystals ``fragile'' and particularly susceptible to structural transitions. The dependence of exponents on the interaction potential arises from the scaling of the overlaps, and  is a well-known feature of jamming that distinguishes it from usual critical phenomena.  By comparing with numerical simulations, we showed that predictions based on the distributions of local areas reproduces  the scaling properties of several global variables remarkably well (Fig. \ref{scaling_of_cycles_fig}). 
Our mean-field description treats the contact triangles as non-interacting entities. Computing the contributions from the correlations between these individual units is non-trivial, and numerical results indicate corrections to the global exponents derived in this Letter \cite{ramola_chakraborty_arxiv_2016}. In future, we plan to explore these non-mean-field effects on the unjamming transition.

{\it Acknowledgements:}
This work has been supported by NSF-DMR 1409093 and the W. M. Keck Foundation.



\clearpage

\begin{widetext}

\begin{appendix}

\section*{\large Supplemental Material for\\ ``Scaling Theory for the Fricitionless Unjamming Transition"}

In this document we provide supplemental figures and details of the calculations presented in the main text.

\section{Distribution of Areas}
In Fig. \ref{area_distribution_fig} we plot the distribution of the normalized areas $\alpha = a_{g,c}/\sigma_g^2$, obtained from numerical simulations, at different energies $E_G$. We simulate a system of bidispersed disks, which causes peaks to occur at five values of $\alpha$ corresponding to the different possible combinations of disks within a $z_v = 3$ (ordered) cycle (see ordered structures section). The peak at $\alpha = 1/2$ corresponds to the ``disordered divergence" $p_{\rm DO}(\alpha)$ whereas the other five peaks correspond to the ``ordered divergences" ($p_{\rm O}(\alpha)$).
These peaks ($p_{\rm DO}(\alpha)$ and $p_{\rm O}(\alpha)$) get sharper as $E_G \to 0^+$ and are infinitely sharp at the transition. The rest of the distribution represents the ``regular part" $p_{\rm reg}(\alpha)$ that does not have a diverging energy dependence as  $E_G \to 0^+$.

\begin{figure}[h!]
\includegraphics[height=0.3\columnwidth,angle=0]{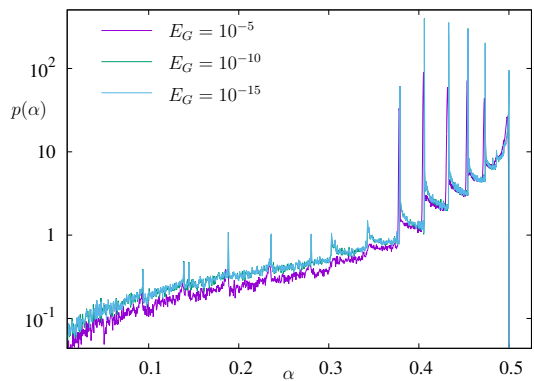}
\caption{The distribution of the normalized areas $\alpha = a_{g,c}/\sigma_g^2$, obtained from numerical simulations, at different energies $E_G$. The plot shows the distribution of $\alpha$ for $N_G = 2048$ bidispersed disks with diameter ratio $1:1.4$ interacting via harmonic potentials ($\mu = 2$). $\alpha \to 1/2$ corresponds to disks with relative contact angles close to $\pi/2$. The peak at $\alpha = 1/2$ corresponds to the ``disordered divergence" $p_{\rm DO}(\alpha)$ whereas the other five peaks correspond to ``ordered divergences" ($p_{\rm O}(\alpha)$) arising from $z_v = 3$ cycles (five different possibilities for a bidispersed system, see ordered structures section). The peaks get sharper as $E_G \to 0^+$ and are infinitely sharp at the transition. The rest of the distribution corresponds to the ``regular part" $p_{\rm reg}(\alpha)$ and does not have a diverging energy dependence as  $E_G \to 0^+$.}
\label{area_distribution_fig}
\end{figure} 

\section{Two point distribution $p(\vec{r}_1, \vec{r}_2)$}
In this section we develop a diagrammatic expansion for the two point distributions of contact vectors $p(\vec{r}_1, \vec{r}_2)$.
From Eq. (10) in the main text we have
\begin{equation}
p(\vec{r}_1,\vec{r}_2) = p(|\vec{r}_1|) p(|\vec{r}_2|) \rho(\theta).
\end{equation}
along with
\begin{equation}
\int d^2 \vec{r}_2 p(\vec{r}_1,\vec{r}_2) = p(\vec{r}_1) = \frac{1}{2 \pi} p(|\vec{r}_1|) ,
\label{one_point_distribution}
\end{equation}
where $p(\vec{r}_1)$ represents the one point distribution of contact vectors.
We then have 
\begin{equation}
\rho(\theta) = \frac{p(\vec{r}_1,\vec{r}_2)}{p(|\vec{r}_1|) p(|\vec{r}_2|)}.
\end{equation}
This function $\rho(\theta)$ therefore encodes the non-trivial correlations between the vectors that arise from the loop constraints. These constraints depend on the number of sides $z_v$ within each cycle.
In order to compute the above function $\rho(\theta)$, it is therefore useful to split the joint distribution of the vectors $\vec{r}_1, \vec{r}_2$ into separate categories based on the minimum cycles to which they belong. We do this as follows
\begin{equation}
p(\vec{r}_1,\vec{r}_2) = \underbrace{3 p(3) p(\vec{r}_1,\vec{r}_2 |3)}_{p(\vec{r}_1,\vec{r}_2 ,3)} + \underbrace{4 p(4) p(\vec{r}_1,\vec{r}_2 |4)}_{p(\vec{r}_1,\vec{r}_2 ,4)} + ...
\end{equation}
where $p(n)$ is the probability of occurrence of a minimum cycle with $z_v = n$ sides, $p(\vec{r}_1,\vec{r}_2 |n)$ is the conditional probability that given a cycle with $z_v = n$ sides two adjacent vectors are $\vec{r}_1,\vec{r}_2$, $p(\vec{r}_1,\vec{r}_2, n)$ represents the joint probability of occurrence of vectors $\vec{r}_1,\vec{r}_2$ together with a cycle of $z_v = n$ sides, and the combinatorial factor accounts for the different ways in which the vectors can be placed within the cycle.

\begin{figure}
\includegraphics[width=0.8\columnwidth,angle=0]{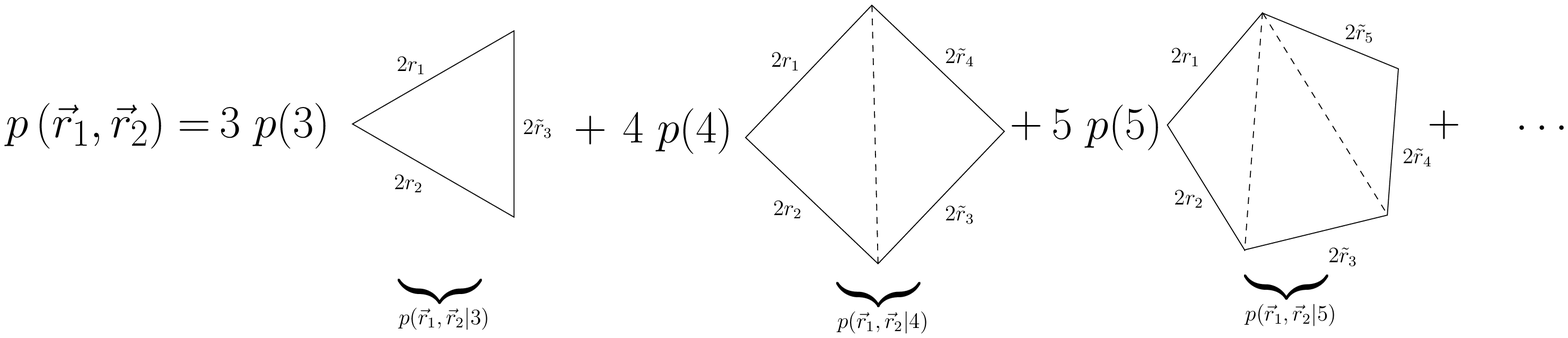}
\caption{Diagrams appearing in the expansion for the joint distribution $p(\vec{r}_1,\vec{r}_2)$. The different terms correspond to the minimum cycles with different numbers of sides. The vectors $\tilde{r}$ are integrated over. The terms corresponding to $z_v = 3$ have a fixed length for all the sides as $E_G \to 0^+$, and therefore give rise to $\rho(\theta)$ localized around a single value $\theta = \arcsin \frac{\sqrt{3}}{2}$. The terms corresponding to $z_v > 3$ have unconstrained sides (depicted with dashed lines) and therefore contribute a finite amount to $\rho(\theta)$ at $\theta = \pi/2$.}
\label{diagrammatic_expansion_figure}
\end{figure} 

Next, the marginal distribution of these two vectors can be computed from the joint distribution of all the vectors in the cycle as
\begin{equation}
p(\vec{r}_1,\vec{r}_2|n) = \int_{0}^{1} d\tilde{r}_3 \int_{0}^{1} d \tilde{r}_4 ...\int_{0}^{1} d\tilde{r}_n p\left( \vec{r}_1, \vec{r}_2, \vec{\tilde{r}}_3, \vec{\tilde{r}}_4, ... \vec{\tilde{r}}_n | n\right),
\end{equation}
where $p\left( \vec{r}_1, \vec{r}_2, \vec{\tilde{r}}_3, \vec{\tilde{r}}_4, ... \vec{\tilde{r}}_n | n\right)$ represents the probability that a given minimum cycle of $z_v = n$ sides has the (ordered) set of vectors $\vec{r}_1, \vec{r}_2, \vec{\tilde{r}}_3, \vec{\tilde{r}}_4, ... \vec{\tilde{r}}_n $. We represent this decomposition as a diagrammatic expansion in Fig \ref{diagrammatic_expansion_figure}.

In order to proceed further, we next make the crucial assumption that the joint probability of occurence of the $n$ vectors can be represented as a product form, along with the loop constraints. We have
\begin{equation}
p\left( \vec{r}_1, \vec{r}_2, \vec{\tilde{r}}_3, \vec{\tilde{r}}_4, ... \vec{\tilde{r}}_n | n \right) = p( \vec{r}_1) p( \vec{r}_2) p(\vec{\tilde{r}}_3)p( \vec{\tilde{r}}_4) ... p(\vec{\tilde{r}}_n) \times \delta \left(\vec{r}_1 + \vec{r}_2 + \vec{\tilde{r}}_3 + \vec{\tilde{r}}_4 + .... \vec{\tilde{r}}_n \right),
\label{product_assumption}
\end{equation} 
where each $p(r)$ is chosen from the one point distribution in Eq. (\ref{one_point_distribution}). This somewhat drastic assumption is justified by the very good agreement between the angular and area distributions obtained from numerical simulations and those obtained by this analysis. This highlights the fact that the crucial correlations in the system arise primarily from these loop constraints.
Finally, in order to simplify the analysis further, we assume that all the disks have the same radii (monodisperse), and that each of the contact vector lengths are drawn from a uniform distribution with width $E_G^{1/\mu}$, consistent with the scaling form provided in Eq. (8) in the main text. We have
\begin{equation}
p\left(r = \frac{|\vec{r}|}{\sigma_g}\right) = \frac{1}{{E_G}^{1/\mu}}\Theta\left(r-1 + {E_G}^{1/\mu}\right)\Theta\left(1-r\right).
\label{uniform_distribution}
\end{equation}
As the energy of the system approaches zero, the fluctuations in the lengths decrease and $r \sim 1$. From Fig. \ref{diagrammatic_expansion_figure} it is clear that there is a fundamental difference between cycles with $z_v = 3$ and $z_v >3$ sides. 
This is because the structures with $z_v = 3$ have a fixed length for all the sides as $E_G \to 0^+$, and therefore give rise to $\rho(\theta)$ localized around a single value $\theta = \arcsin \frac{\sqrt{3}}{2}$. The terms corresponding to $z_v > 3$ have unconstrained sides (as depicted with dashed lines) and therefore contribute a finite amount to $\rho(\theta)$ at $\theta = \pi/2$. We can also explicitly derive the distribution of angles ($p(\sin \theta)$) for the $z_v = 3$ case using the above assumptions, which we detail in the next section.
 
\subsection{Ordered Structures ($z_v = 3$)}

\begin{figure}
\includegraphics[width=0.45\columnwidth,angle=0]{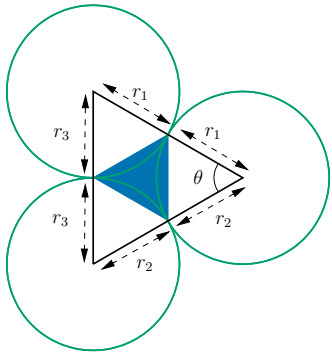}
\caption{The three-disk minimum cycle ($z_v = 3$) used to compute the the area distribution in Eq. (\ref{ordered_integral_area}) and angular distribution in Eq. (\ref{ordered_integral_theta}). In our analysis we focus only on the case where all the disks have an equal size.}
\label{ordered_expansion_figure}
\end{figure} 

\begin{figure}
\includegraphics[width=0.4\columnwidth,angle=0]{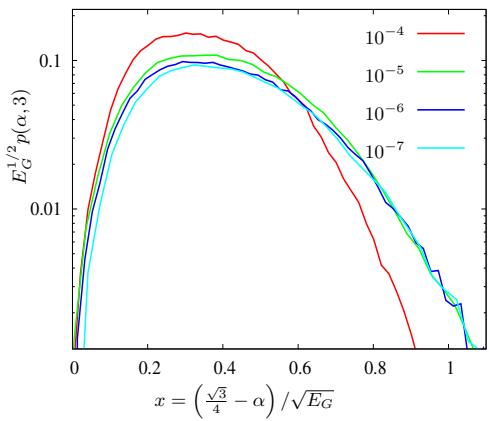}
\caption{Scaling collapse of the distribution of areas of the ordered ($z_v = 3$) cycles $p(\alpha,3)$ obtained from numerical simulations, around the divergence at $\alpha = \sqrt{3}/4$. The plot shows the distribution of $\alpha$ for $N_G = 4096$ disks interacting via harmonic potentials ($\mu = 2$) at different energies. The scaling is consistent with Eq. (4) in the main text.}
\label{scaling_collapse_order_fig}
\end{figure} 

In this section, we compute the distribution of areas for the $z_v = 3$ cycles and provide the scaling form for the ``ordered divergence" mentioned in the main text.
From Fig. \ref{ordered_expansion_figure} it is straightforward to compute
\begin{equation}
\sin \theta = \frac{\sqrt{4 r_1^2 r_2^2 - \left( r_1^2 + r_2^2 - r_3^2\right)}}{2 r_1 r_2}.
\label{heron_theta}
\end{equation}
Using the above expression, the distribution of areas for the $z_v = 3$ cycles can be computed as
\begin{equation}
p(\alpha,3) = \int_{0}^{1} d r_1 \int_{0}^{1} d r_2 \int_{0}^{1} d r_3
~p(r_1) p(r_2) p(r_3) ~\delta \left(\alpha - \frac{\sqrt{4 r_1^2 r_2^2 - \left( r_1^2 + r_2^2 - r_3^2\right)}}{4} \right).
\label{ordered_integral_area}
\end{equation}
Next, replacing $p(r)$ with the uniform distributions in Eq. (\ref{uniform_distribution}) leads to the following scaling form for the ordered divergence in the distribution of areas
\begin{equation}
p(\alpha,3) = \frac{1}{{E_G}^{1/\mu}}\mathcal{P}_{\rm O}\left(\frac{\alpha - \frac{\sqrt{3}}{4}}{{E_G}^{1/\mu}}\right),
\end{equation}
which is Eq. (4) in the main text.
In Fig. \ref{scaling_collapse_order_fig} we plot the scaling collapse of the distribution $p(\alpha,3)$ obtained from numerical simulations, around the divergence at $\alpha = \sqrt{3}/4$. The scaling is consistent with the above analysis and with Eq. (4) in the main text.
The scaling behaviour of the distribution obtained from our theoretical analysis is illustrated in Fig. \ref{ordered_comparision_fig} where we plot the distribution of areas computed numerically using Eq. (\ref{ordered_integral_area}), along with the distributions obtained from numerical simulations. We find a good agreement between the distributions in the limit $\alpha \to \sqrt{3}/4$. The scaling function has the following behaviour
\begin{equation}
\mathcal{P}_{\rm O}(x) \sim x^2  ~~~{\rm for}~~~ x\to 0.
\end{equation}
We note that in our analysis we have only focussed on monodispersed disks. The generalization to the polydisperse case involves all combinations of disks that can produce a $z_v = 3$ cycle. For the bidispersed case with diameter ratio $1:1.4$ that we simulate, the peaks in the area distribution occur at $\alpha = \sqrt{3}/4 = 0.433013$ (for equal sized disks), $0.406116, 0.45453, 0.378775$ and $0.473803$ \cite{ramola_chakraborty_arxiv_2016} (see Fig. \ref{area_distribution_fig}). The scaling analysis for each of these cases remains the same.

Similarly, using the product assumption in Eq. (\ref{product_assumption}), the distribution of the angles $p(\sin \theta)$ for finite energies corresponding to the $z_v = 3$ cycles can be explicitly computed as
\begin{equation}
p(\sin \theta,3) = \int_{0}^{1} d r_1 \int_{0}^{1} d r_2 \int_{0}^{1} d r_3  
~p(r_1) p(r_2) p(r_3) ~\delta \left( \sin \theta - \frac{\sqrt{4 r_1^2 r_2^2 - \left( r_1^2 + r_2^2 - r_3^2\right)}}{2 r_1 r_2} \right).
\label{ordered_integral_theta}
\end{equation}
Next, replacing $p(r)$ with the uniform distributions in Eq. (\ref{uniform_distribution}) we find the following scaling form for the angular distribution
\begin{equation}
p(\sin \theta,3) = \frac{1}{{E_G}^{1/\mu}}\mathcal{P}_{\theta}\left(\frac{\sin \theta - \frac{\sqrt{3}}{2}}{{E_G}^{1/\mu}}\right).
\label{scaling_form_theta}
\end{equation}
This behaviour is illustrated in Fig. \ref{ordered_comparision_fig} where we plot the angular distributions computed numerically using Eq. (\ref{ordered_integral_theta}) for $\mu =2$.

\begin{figure}
\includegraphics[height=0.3\columnwidth,angle=0]{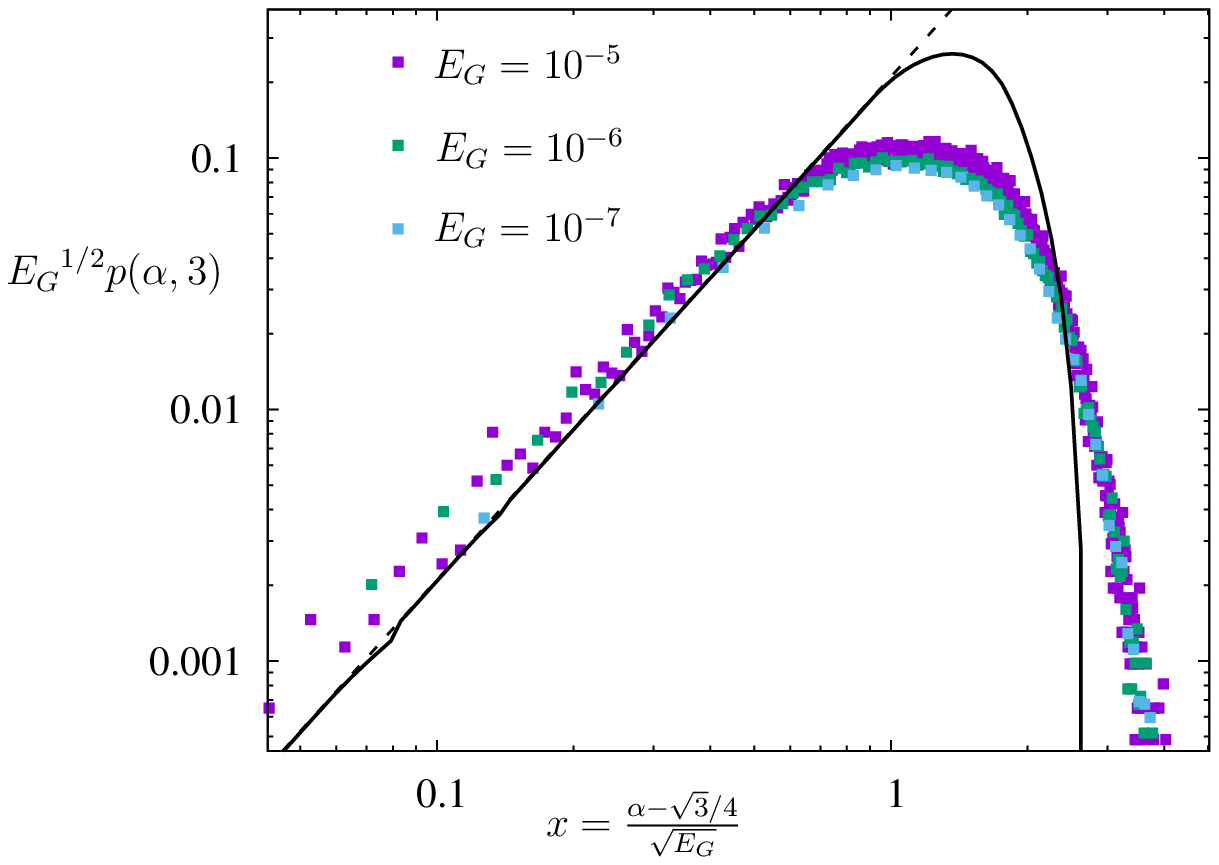}
\includegraphics[height=0.3\columnwidth,angle=0]{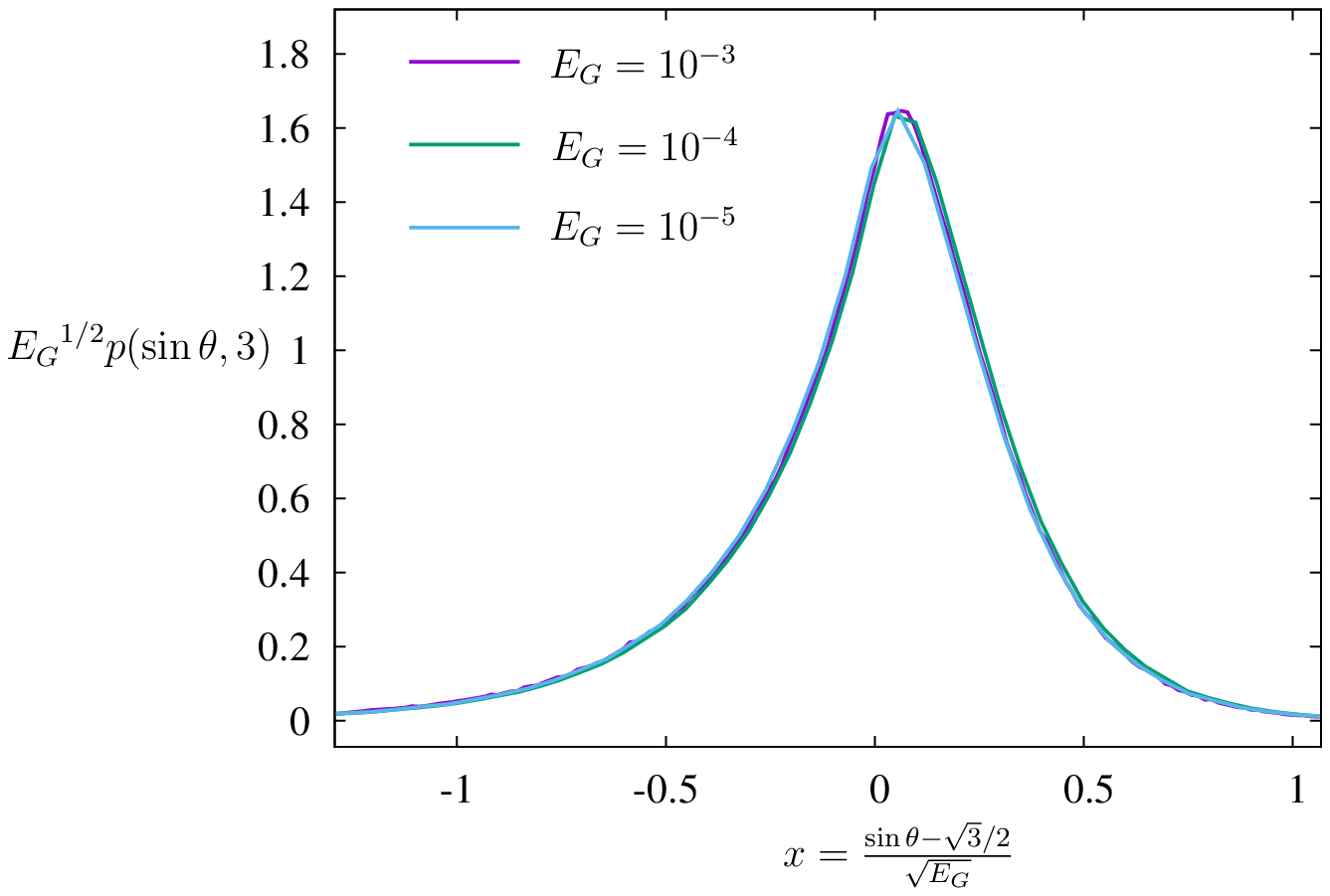}
\caption{{\bf (Left)} Scaling collapse of the distribution of contact areas of the $z_v = 3$ cycles obtained from numerical simulations along with the prediction from the theory. The plot shows the distribution of $\alpha$ for $N_G = 4096$ bidispersed disks interacting via harmonic potentials ($\mu = 2$). The bold line represents the theoretical distribution obtained by numerically integrating Eq. (\ref{ordered_integral_area}) using a uniform distribution of contact vector lengths given in Eq. (\ref{uniform_distribution}) with $\mu = 2$. The dashed line has a slope $2$. {\bf (Right)} Scaling collapse of the theoretical distribution of $p(\sin \theta, 3)$ obtained by numerically integrating Eq. (\ref{ordered_integral_theta}) for $\mu = 2$. The distribution obeys the scaling form provided in Eq. (\ref{scaling_form_theta}).}
\label{ordered_comparision_fig}
\end{figure} 

\section*{Disordered Divergence: $p_{\rm DO}(\alpha)$ }
In this section we derive an expression for the disordered divergence $p_{\rm DO}(\alpha)$ in the distribution of areas.
We begin by assuming a product form for the joint distribution of contact vectors
\begin{equation}
p(\vec{r}_1,\vec{r}_2) = \frac{1}{2 \pi} p(r_1) p(r_2).
\end{equation}
In the above decomposition, we have assumed a uniform distribution for $\rho(\theta)$ in the region $[0,2 \pi]$. This assumption is justified since we are interested in the distribution close to $\theta \sim \pi/2$. The analysis presented in this section can easily be generalized to smaller ranges of $\theta$.  We have checked that the scaling features of the distribution near the transition are unchanged by extending the range of $\theta$. We then have
\begin{equation}
\rho(\sin \theta) = \frac{1}{2 \pi} \frac{1}{\sqrt{1 - \sin^2 \theta}}.
\end{equation}

Next, from Eq. (9) in the main text we have the following equation for the distribution of the areas
\begin{equation}
p(\alpha) = \int_{0}^{1} d r_1 \int_{0}^{1} d r_2 \int_{0}^{1} d \sin \theta 
~p(r_1) p(r_2) \rho(\sin \theta) ~\delta \left( \frac{1}{2} r_1 r_2 \sin \theta - \alpha \right).
\end{equation}
Once again to simplify the analysis, we replace the one point distribution of contact vector lengths $p(r)$ by the uniform distribution in Eq. (\ref{uniform_distribution}).
Finally, performing the integral over $\sin \theta$ we arrive at the following expression for the distribution of areas
\begin{equation}
p_{\rm DO}\left(\alpha\right) = \frac{4}{\pi {E_G}^{2/\mu}}\int_{1-{E_G}^{1/\mu}}^{1} \int_{1-{E_G}^{1/\mu}}^{1} \frac{\Theta(xy - 2 \alpha)}{\sqrt{x^2 y^2-4 \alpha^2}} dx dy.
\end{equation}
In order to perform this computation we compute the simpler indefinite integral defined as
\begin{equation}
S_{\rm DO}(\alpha, x,y) = \int \int \frac{1}{\sqrt{x^2 y^2-4 \alpha^2}} dx dy.
\end{equation}
This does not explicitly contain the $\Theta$ function. We can account for the  $\Theta(xy - 2 \alpha)$ constraint by breaking the definite integral into regions depending on the value of $\alpha$.
The definite integral can then be expressed as combinations of the above indefinite integral. We have
\begin{equation}
p_{\rm DO}(\alpha) = \frac{4}{\pi {E_G}^{2/\mu}} \left( S_{\rm DO}(\alpha,1,1) -  S_{\rm DO}(\alpha,1,1-{E_G}^{1/\mu}) - S_{\rm DO}(\alpha,1-{E_G}^{1/\mu},1) + S_{\rm DO}(\alpha,1-{E_G}^{1/\mu},1-{E_G}^{1/\mu}) \right).
\label{final_expression}
\end{equation}

\begin{figure}
\includegraphics[height=0.4\columnwidth,angle=0]{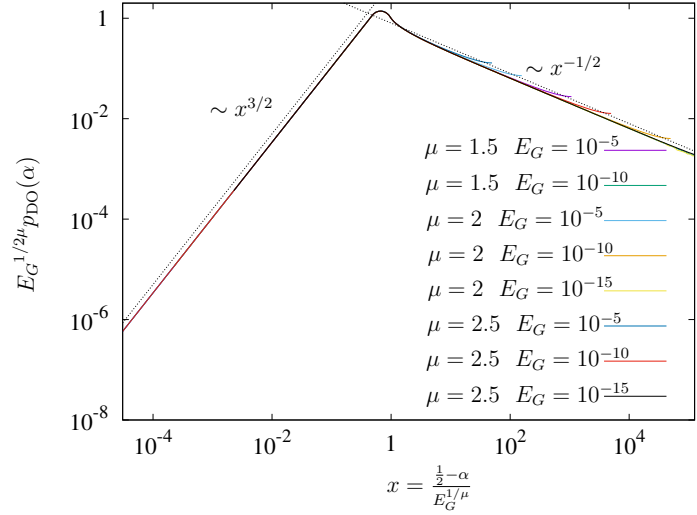}
\caption{Scaling collapse of the theoretical distribution of $p_{\rm DO}(\alpha)$ obtained from Eqs. (\ref{final_expression}) and (\ref{exact_expression_disorder}) for different repulsive potentials $\mu = 1.5,2$ and $2.5$ at varying global energies. The distribution obeys the scaling form provided in Eq. (2) in the main text. The scaling function has the limiting behaviours announced in Eq. (3) in the main text.}
\label{pdo_collapse_different_mu_fig}
\end{figure} 

\subsection{Explicit expression for $S_{\rm DO}(\alpha,x,y)$}
We derive below an exact expression for the above indefinite integral $S_{\rm DO}(\alpha, x,y)$.
First, performing the integral over $x$ we arrive at
\begin{equation}
S_{\rm DO}(\alpha,x,y) =  \int  \frac{\log \left(y \sqrt{x^2 y^2-4 \alpha^2}+x y^2\right)}{y} dy.
\end{equation}
Next, the integral with respect to $y$ can be performed exactly. After some algebraic simplifications (using Mathematica), the explicit expression is
\begin{eqnarray}
\nonumber
S_{\rm DO}(\alpha,x,y) &=& -\frac{1}{2} \text{Li}_2\left(\frac{2 \alpha^2}{2 \alpha^2-x y \left(x y+\sqrt{x^2 y^2-4
   \alpha^2}\right)}\right)-\frac{1}{2} \log ^2\left(\sqrt{x^2 y^2-4 \alpha^2}+x y\right)+ \log (x y) \log
   \left(\sqrt{x^2 y^2-4 \alpha^2}+x y\right)\\
   & + &  \log (2) \log \left(\frac{\sqrt{x^2 y^2-4
   \alpha^2}}{x}+y\right)+\frac{1}{2} \log (\alpha) \log \left(\frac{\alpha}{x^2}\right)+\frac{1}{2} \log ^2(y)-\frac{\pi ^2}{8}-\frac{1}{2}
   \log ^2(2),
   \label{exact_expression_disorder}
\end{eqnarray}
where $\text{Li}_2$ is the Polylogarithm function.
Although the above expression is not explicitly symmetric under the $(x,y) \to (y,x)$ transformation, it is easy to see that the expression $p_{\rm DO}(\alpha)$ preserves this symmetry.
Using Eq. (\ref{exact_expression_disorder}) it is straightforward to show (for example, using Mathematica) that the function $p_{\rm DO}(\alpha)$ given in Eq. (\ref{final_expression}) has the asymptotic behaviours mentioned in the scaling form in Eq. (3) in the main text.

In Fig. \ref{pdo_collapse_different_mu_fig} we show the scaling collapse of the theoretical distribution $p_{\rm DO}(\alpha)$ obtained from Eqs. (\ref{final_expression}) and (\ref{exact_expression_disorder}) for different repulsive potentials $\mu = 1.5,2$ and $2.5$. The distribution obeys the scaling form provided in Eq. (2) in the main text. The scaling function has the limiting behaviours announced in Eq. (3) in the main text.


\end{appendix}
\clearpage

\end{widetext}

\end{document}